\documentclass[a4paper,11pt]{article}
\usepackage{jcappub}
\usepackage[utf8]{inputenc}
\usepackage{amsmath,amsfonts,amssymb}
\usepackage{graphicx}
\usepackage[toc,page]{appendix}
\usepackage{cleveref}
\usepackage{booktabs}
\usepackage{mathtools}
\usepackage{microtype}
\usepackage{subcaption}
\usepackage{comment}
\newcommand{\mpl}{M_{\rm Pl}}
\newcommand{\dd}{\mathrm{d}}

\newcommand{\Pzeta}{\mathcal{P}_{\zeta}}

\begin{document}

\title{Primordial black holes and gravitational waves
from aligned axion inflation}

\author[a,b]{Nadir Ijaz}

\affiliation[a]{Dipartimento di Fisica e Astronomia “G. Galilei”, Universit\`a degli Studi di Padova, via Marzolo 8, I-35131 Padova, Italy}
\affiliation[b]{INFN, Sezione di Padova, via Marzolo 8, I-35131 Padova, Italy}

\emailAdd{nadir.ijaz@phd.unipd.it}

\abstract{We study the production of primordial black holes (PBHs) and scalar-induced gravitational waves (SIGWs) in aligned axion inflation through the transition between two stages of inflation. The first stage proceeds along a metastable trajectory and ends when this trajectory loses stability, causing the fields to turn towards the attractor of the second inflationary stage. During the transition, coupled adiabatic--entropy evolution produces a localized enhancement of the curvature power spectrum. We compute the PBH abundance from this spectrum assuming Gaussian primordial curvature perturbations and using the nonlinear compaction-function formalism with a spectrum-dependent collapse threshold and critical collapse. Representative benchmark solutions yield asteroid-mass PBHs with a potentially significant dark-matter abundance, together with an associated SIGW background.}

\maketitle

\section{Introduction}
\label{sec:introduction}

Inflation provides a compelling framework for the physics of the early Universe~\cite{Guth:1980zm,Linde:1981mu,Albrecht:1982wi}. Besides addressing the horizon, flatness, and relic problems of the standard hot Big Bang cosmology, it provides a mechanism through which quantum fluctuations are stretched to cosmological scales and seed the primordial inhomogeneities that later develop into the observed cosmic structure~\cite{Mukhanov:1981xt,Hawking:1982cz,Starobinsky:1982ee,Guth:1982ec, Bardeen:1983qw}. Measurements of the cosmic microwave background (CMB) are consistent with predominantly adiabatic, nearly Gaussian perturbations with an approximately scale-invariant spectrum, and place increasingly stringent constraints on the scalar spectral index $n_s$ and the tensor-to-scalar ratio $r$~\cite{Planck:2018jri,BICEP:2021xfz}. Nevertheless, the nature of the degrees of freedom responsible for inflation, and their connection to a fundamental theory of particle physics, remain unresolved.

A successful scalar-field realization of inflation requires the potential to remain sufficiently flat over the portion of field space relevant for the accelerated expansion. This condition is especially non-trivial in large-field models, in which the inflaton traverses a distance comparable to or larger than the reduced Planck mass, $\mpl\equiv(8\pi G)^{-1/2}$. Such models are particularly interesting because a sufficiently large field excursion can be associated with an observable primordial tensor signal~\cite{Lyth:1996im}. At the same time, higher-dimensional operators suppressed by the ultraviolet cutoff can easily distort the potential and spoil slow roll. Maintaining a controlled inflationary potential therefore generally requires either a hierarchy among the coefficients of these operators or a symmetry that protects the flat direction.

Axion-like fields provide a natural setting in which such protection can be realized. Their continuous shift symmetry is broken nonperturbatively to a discrete subgroup, generating a periodic potential while suppressing perturbative corrections. The simplest realization is natural inflation, in which a single axion evolves in a cosine potential~\cite{Freese:1990rb,Adams:1992bn,Pajer:2013fsa}. Agreement with the observed scalar tilt generally requires a super-Planckian effective axion decay constant, raising questions about the validity of the effective field-theory description~\cite{Banks:2003sx}. In particular, continuous global symmetries are not expected to remain exact in quantum gravity. Gravitational instantons and wormholes may therefore break the axionic shift symmetry and generate additional contributions to the potential~\cite{Ghigna:1992iv,Giddings:1987cg,Giddings:1989bq,Holman:1992us,Kamionkowski:1992mf,Kallosh:1995hi}.

Several mechanisms have been proposed to realize a parametrically large inflationary field range while keeping the fundamental axion decay constants below $M_{\rm Pl}$. These include the collective dynamics of many axions in N-flation~\cite{Dimopoulos:2005ac}, the unwrapping of the axion periodicity through monodromy~\cite{McAllister:2008hb}, and the alignment mechanism~\cite{Kim:2004rp}. In the latter construction, two periodic contributions to the potential are characterized by nearly aligned inverse-decay-constant vectors. Close to the alignment limit, one linear combination of the axions remains heavy, whereas the orthogonal combination develops a parametrically shallow potential and behaves as an effective axion with an enhanced decay constant. The compatibility of the alignment mechanism with quantum gravity and the weak gravity conjecture has been extensively investigated~\cite{Arkani-Hamed:2006emk,Rudelius:2015xta,Montero:2015ofa,Heidenreich:2015wga}. Its possible realizations, as well as the associated limitations, in string compactifications have also been studied~\cite{Long:2014dta,Palti:2015xra,Gao:2014uha,Hebecker:2018fln,Angus:2021jpr}.

The complete two-field dynamics of aligned axion inflation is richer than the effective single-field description suggests. In addition to stable trajectories that evolve continuously towards a minimum of the potential, the model admits metastable inflationary trajectories located close to saddle points~\cite{Peloso:2015dsa}. Along such solutions, the shallow direction supports slow-roll inflation, while the orthogonal direction initially stabilizes the background. Inflation on this branch ends when the transverse stabilization is lost and the trajectory becomes unstable. Since most of the first inflationary stage takes place near a relatively flat saddle region, these metastable solutions predict a tensor-to-scalar ratio smaller than that of the stable branch and can remain compatible with CMB constraints. The analytic construction of the metastable trajectory, the condition determining its endpoint, and the corresponding leading slow-roll predictions were derived in Ref.~\cite{Greco:2024ngr}. The loss of transverse stability does not generally bring the system directly to a minimum. The fields can instead undergo a short non-adiabatic transition, involving a substantial displacement in field space and a sequence of turns, before approaching a neighbouring valley connected to the vacuum. A second stage of inflation may then take place along this valley~\cite{Greco:2024ngr}.

The transition between the two stages of inflation involving sharp turns in the field space requires a complete multifield treatment. Similar transitions and sharp turns in multifield models have been shown to generate significant localized enhancements of the curvature power spectrum~\cite{Braglia:2020eai,Palma:2020ejf,Fumagalli:2020adf,Dalianis:2026tty,Boutivas:2022qfh,Christodoulidis:2023eiw,Braglia:2020taf,Braglia:2022phb,Ijaz:2023cvc}. When the corresponding modes re-enter the Hubble horizon during radiation domination, sufficiently large perturbations can collapse and form primordial black holes (PBHs)~\cite{Hawking:1971ei,Carr:1974nx,carr1975primordial}. The same enhanced scalar perturbations source tensor modes at second order, generating a scalar-induced gravitational-wave (SIGW) background~\cite{Ananda:2006af,Baumann:2007zm,Kohri:2018awv}. The phenomenology of PBHs and SIGWs associated with enhanced primordial fluctuations has since been explored extensively; for comprehensive reviews and further references, see Refs.~\cite{Sasaki:2018dmp,Domenech:2021ztg}. PBHs and SIGWs therefore provide complementary probes of inflationary dynamics on scales much smaller than those accessible through the CMB.

Related small-scale signatures have also been studied in axion inflation with an additional gauge sector~\cite{Barnaby:2011qe,Figueroa:2023oxc,Garcia-Bellido:2023ser,Barbon:2025wjl}. Gauge quanta amplified by the motion of the axion can source scalar perturbations sufficiently large to produce PBHs~\cite{Linde:2012bt,Bugaev:2013fya}. The accompanying stochastic gravitational-wave signatures, arising either directly from the amplified gauge fields or at second order from the sourced scalar perturbations, have been investigated together with PBH production in Refs.~\cite{Garcia-Bellido:2016dkw,Ozsoy:2020kat,Franciolini:2026cps,Cheng:2018yyr,Almeida:2020kaq,Unal:2023srk,Dimastrogiovanni:2024xvc}. A gauge-field extension of aligned axion inflation was recently investigated in Ref.~\cite{Greco:2025nvv}. In that study, the axions were coupled to a $U(1)$ gauge field, and the production of gauge quanta during the transient phase was analyzed. After accounting for the homogeneous backreaction of the produced gauge field on the axion background, the analysis yielded a strongly scale-dependent spectrum of gauge-sourced gravitational waves, whose characteristic scale is determined by the duration of the second inflationary stage.

In the present work, we extend the phenomenological scope of the aligned axion model by investigating whether its intrinsic multifield dynamics can generate observable small-scale signatures without an additional gauge sector. We solve the coupled adiabatic and entropy perturbation
equations through the loss of transverse stability and the
subsequent turns towards the second inflationary valley. This coupled dynamics generates a localized enhancement of the curvature power spectrum on small scales, while leaving the CMB-scale spectrum essentially unchanged. Using the enhanced spectrum, we compute the PBH abundance within the nonlinear compaction-function formalism, determining the collapse threshold from the shape of the spectrum and accounting for the critical scaling of the PBH mass near the threshold. We also calculate the associated SIGW background sourced by the same scalar perturbations. For representative benchmark points, this intrinsic mechanism yields an asteroid-mass population of PBHs with a potentially significant dark-matter abundance, accompanied by a SIGW signal.

The paper is organized as follows. In Sec.~\ref{sec:model}, we introduce the aligned axion inflation model and summarize the properties of the metastable inflationary solutions. In Sec.~\ref{sec:background-dynamics}, we discuss the homogeneous two-field dynamics and characterize the instability and the transition between the two inflationary stages. In Sec.~\ref{sec:perturbations}, we formulate and solve the coupled adiabatic and entropy perturbation equations and present the resulting curvature spectrum. In Sec.~\ref{sec:pbh}, we study the formation, mass distribution, and present-day abundance of primordial black holes. In Sec.~\ref{sec:sigw}, we present the associated SIGW spectra. We summarize our results in Sec.~\ref{sec:conclusion}. Technical details of the PBH calculation are collected in the appendix.

\section{The aligned axion model}
\label{sec:model}

We consider two canonically normalized axions, $\boldsymbol\phi=(\phi_1,\phi_2)$, with components $\phi^I$, minimally coupled to gravity,
\begin{equation}
S =\int d^4x\,\sqrt{-g}\left[\frac{\mpl^2}{2}R -\frac{1}{2}\delta_{IJ} \partial_\mu\phi^I\partial^\mu\phi^J -V(\phi_1,\phi_2) \right],
\label{eq:model-action}
\end{equation}
with the aligned potential~\cite{Kim:2004rp}
\begin{equation}
V(\phi_1,\phi_2) = \Lambda_1^4 \left[1-\cos\!\left(\boldsymbol q_1\cdot\boldsymbol\phi\right)\right] + \Lambda_2^4 \left[1-\cos\!\left(\boldsymbol q_2\cdot\boldsymbol\phi\right)\right],
\label{eq:model-potential-original}
\end{equation}
where
\begin{equation}
\boldsymbol q_i = \left(\frac{1}{f_i},\frac{1}{g_i}\right), \qquad i=1,2,
\label{eq:q-vectors}
\end{equation}
are the inverse-decay-constant vectors that determine the two periodic directions in field space. We order the two contributions such that $\Lambda_1\geq\Lambda_2$, and define
\begin{equation}
\Lambda^4\equiv\Lambda_1^4+\Lambda_2^4, \qquad r_\Lambda\equiv\frac{\Lambda_2^4}{\Lambda_1^4}, \qquad 0<r_\Lambda\leq1.
\label{eq:potential-scales}
\end{equation}
The parameter $\Lambda$ fixes the overall inflationary scale, whereas $r_\Lambda$, together with the relative orientation and magnitudes of the vectors $\boldsymbol q_i$, determines the geometry of the potential.

The canonical kinetic term is invariant under a constant rotation of the two fields. It is therefore convenient to characterize the potential through quantities that do not depend on the chosen field basis. We use
\begin{equation}
n_1\equiv\boldsymbol q_1^2 =\frac{1}{f_1^2}+\frac{1}{g_1^2}, \qquad
n_2\equiv\boldsymbol q_2^2 =\frac{1}{f_2^2}+\frac{1}{g_2^2},
\label{eq:q-norms}
\end{equation}
and
\begin{equation}
{\cal C} \equiv \frac{1}{f_2g_1}-\frac{1}{f_1g_2}.
\label{eq:q-determinant}
\end{equation}

The quantities $n_i$ are the squared lengths of the two inverse-decay-constant vectors, while ${\cal C}$ measures their oriented area, with the sign convention in Eq.~\eqref{eq:q-determinant}. They satisfy
\begin{equation}
{\cal C}^2 = n_1n_2-(\boldsymbol q_1\cdot\boldsymbol q_2)^2.
\label{eq:alignment-geometry}
\end{equation}
These combinations remove the redundancy associated with the orientation of the canonical field axes and characterize the periodicities and their relative alignment~\cite{Greco:2024ngr}. Exact alignment corresponds to ${\cal C}=0$, for which the two vectors are parallel and the orthogonal field direction is absent from the potential. Following Ref.~\cite{Greco:2024ngr}, we quantify the departure from exact alignment by the dimensionless parameter
\begin{equation}
\gamma\equiv\frac{2{\cal C}}{n_1+n_2}, \qquad |\gamma|\ll1.
\label{eq:alignment-parameter}
\end{equation}

For the numerical implementation, we choose the basis in which $g_2^{-1}=0$. A constant orthogonal rotation always allows this choice without changing the canonical kinetic terms. With the signs chosen such that $f_2^{-1}>0$ and $\boldsymbol q_1\cdot\boldsymbol q_2>0$, using the evenness of the cosines, the inverse-decay-constant vectors are
\begin{equation}
\boldsymbol q_2=(\sqrt{n_2},0), \qquad \boldsymbol q_1= \left( \sqrt{n_1-\frac{{\cal C}^2}{n_2}}, \frac{{\cal C}}{\sqrt{n_2}} \right), \qquad {\cal C}=\frac{\gamma}{2}(n_1+n_2).
\label{eq:numerical-basis}
\end{equation}
These are the same vectors defined in Eq.~\eqref{eq:q-vectors}, written explicitly in the chosen basis.

For the study of the metastable branch, we place the relevant saddle at the origin by defining the shifted fields
\begin{equation}
\theta=\phi_1, \qquad \rho=\phi_2-\frac{\pi\sqrt{n_2}}{{\cal C}}, \qquad {\cal C}\neq0.
\label{eq:saddle-translation}
\end{equation}
This constant translation shifts the first cosine phase by $\pi$ and leaves the second unchanged:
\begin{equation}
\boldsymbol q_1\cdot\boldsymbol\phi={\cal A}_1+\pi, \qquad \boldsymbol q_2\cdot\boldsymbol\phi={\cal A}_2.
\label{eq:translated-phases}
\end{equation}
The kinetic terms and the vectors $\boldsymbol q_i$ remain unchanged. In terms of $(\theta,\rho)$, the complete Lagrangian density becomes
\begin{equation}
\frac{\mathcal L}{\sqrt{-g}}=\frac{\mpl^2}{2}R-\frac12(\partial\theta)^2-\frac12(\partial\rho)^2 -V(\theta,\rho),
\label{eq:final-lagrangian}
\end{equation}
with
\begin{align}
V(\theta,\rho)
&= \frac{\Lambda^4}{1+r_\Lambda} \left[ 1+r_\Lambda+\cos{\cal A}_1-r_\Lambda\cos{\cal A}_2 \right],
\label{eq:shifted-potential}\\
{\cal A}_1
&= \sqrt{n_1-\frac{{\cal C}^2}{n_2}}\,\theta +\frac{{\cal C}}{\sqrt{n_2}}\,\rho, \qquad {\cal A}_2=\sqrt{n_2}\,\theta.
\label{eq:final-phases}
\end{align}
We use the shifted fields $(\theta,\rho)$ throughout the remainder of this work.

The origin ${\cal A}_1={\cal A}_2=0$ is the saddle conventionally denoted by ${\cal S}_B$. Since the vectors $\boldsymbol q_1$ and $\boldsymbol q_2$ are linearly independent for ${\cal C}\neq0$, all stationary points occur when the two phases are integer multiples of $\pi$. Their complete classification is given in Refs.~\cite{Peloso:2015dsa,Greco:2024ngr}; here we summarize only the properties relevant to the inflationary solutions used in our analysis.

At ${\cal S}_B$, the potential Hessian is
\begin{equation}
\left.V_{,IJ}\right|_{{\cal S}_B}= \frac{\Lambda^4}{1+r_\Lambda} \left(r_\Lambda q_{2I}q_{2J}-q_{1I}q_{1J} \right),
\label{eq:saddle-hessian}
\end{equation}
with eigenvalues
\begin{equation}
m_\pm^2\big|_{{\cal S}_B} = \frac{\Lambda^4}{2(1+r_\Lambda)} \left[r_\Lambda n_2-n_1 \pm \sqrt{\left(r_\Lambda n_2-n_1\right)^2+4r_\Lambda{\cal C}^2 } \right].
\label{eq:SB-eigenmasses}
\end{equation}
For ${\cal C}\neq0$, the two eigenvalues have opposite signs. In the parameter region $r_\Lambda n_2>n_1$, relevant for trajectories departing from ${\cal S}_B$, their strong-alignment expansions, away from the degenerate limit $r_\Lambda n_2=n_1$, are
\begin{align}
m_{\rm heavy}^2\big|_{{\cal S}_B} \equiv m_+^2\big|_{{\cal S}_B}
&= \frac{\Lambda^4}{1+r_\Lambda} \left(r_\Lambda n_2-n_1\right)+\mathcal O({\cal C}^2), \nonumber\\
m_{\rm light}^2\big|_{{\cal S}_B} \equiv m_-^2\big|_{{\cal S}_B} 
&= -\frac{r_\Lambda\Lambda^4{\cal C}^2} {(1+r_\Lambda)(r_\Lambda n_2-n_1)} +\mathcal O({\cal C}^4).
\label{eq:SB-aligned-eigenmasses}
\end{align}
Here, ``heavy'' and ``light'' denote squared eigenmasses of order $\mathcal O({\cal C}^{0})$ and $\mathcal O({\cal C}^{2})$, respectively, in the alignment limit. The inflationary trajectory therefore proceeds along the light tachyonic direction, which becomes flat in the limit of perfect alignment, while the large positive eigenmass stabilizes the orthogonal direction.

The local eigenmass hierarchy at ${\cal S}_B$ requires $n_1/n_2<r_\Lambda$. In the strong-alignment limit, the additional requirement that the orthogonal direction become unstable along the trajectory restricts the metastable domain to~\cite{Greco:2024ngr}
\begin{equation}
r_\Lambda^2<\frac{n_1}{n_2}<r_\Lambda.
\label{eq:metastable-domain}
\end{equation}
Within this domain, the fields undergo prolonged slow-roll evolution along the light direction until transverse stabilization is lost. Such trajectories are therefore termed metastable; they were first identified in Ref.~\cite{Peloso:2015dsa} and subsequently analyzed in detail in Ref.~\cite{Greco:2024ngr}.

Building on these results, we focus on CMB-compatible metastable solutions for which the post-instability dynamics redirect the fields into an adjacent valley. We determine the full two-field background evolution numerically and evolve the coupled scalar perturbations continuously from the metastable phase, through the loss of transverse stability and the ensuing transition, into the second inflationary stage.

\section{Background dynamics}
\label{sec:background-dynamics}

We consider spatially homogeneous scalar fields $\theta(t)$ and $\rho(t)$ evolving in a spatially flat FLRW background,
\begin{equation}
    ds^2=-dt^2+a^2(t)d\boldsymbol{x}^2.
    \label{eq:FLRW-metric}
\end{equation}
The fields $\theta$ and $\rho$ have canonical kinetic terms and a flat field-space metric $\delta_{IJ}$, which is used to raise and lower field-space indices.

The homogeneous field equations are
\begin{align}
    \ddot\theta+3H\dot\theta+V_{,\theta}&=0,
    \label{eq:theta-eom}\\
    \ddot\rho+3H\dot\rho+V_{,\rho}&=0,
    \label{eq:rho-eom}
\end{align}
together with the Friedmann equations
\begin{align}
    3\mpl^2H^2
    &=
    \frac12\left(\dot\theta^2+\dot\rho^2\right)+V,
    \label{eq:friedmann}\\
    \dot H
    &=
    -\frac{\dot\theta^2+\dot\rho^2}{2\mpl^2},
    \label{eq:Hdot}
\end{align}
where overdots denote cosmic-time derivatives. Inflation corresponds to $\epsilon_H<1$, with
\begin{equation}
    \epsilon_H
    \equiv
    -\frac{\dot H}{H^2}
    =
    \frac{\dot\theta^2+\dot\rho^2}{2\mpl^2H^2}.
    \label{eq:epsilon-w}
\end{equation}

To describe the trajectory and its perturbations, we introduce the kinematic adiabatic--entropy basis~\cite{Gordon:2000hv,GrootNibbelink:2001qt}. The speed along the trajectory is
\begin{equation}
    \dot\sigma
    \equiv
    \sqrt{\dot\theta^2+\dot\rho^2},
    \label{eq:sigmadot}
\end{equation}
and the tangent and normal unit vectors are
\begin{equation}
    e_\sigma^I
    =
    \frac{1}{\dot\sigma}(\dot\theta,\dot\rho),
    \qquad
    e_s^I
    =
    \frac{1}{\dot\sigma}(-\dot\rho,\dot\theta).
    \label{eq:kinematic-basis}
\end{equation}
The corresponding projections of the potential gradient and Hessian are
\begin{equation}
    V_a=e_a^I V_{,I},
    \qquad
    V_{ab}=e_a^I e_b^J V_{,IJ},
    \qquad
    a,b\in\{\sigma,s\}.
    \label{eq:potential-projections}
\end{equation}
Projecting the background equations gives
\begin{align}
    \ddot\sigma+3H\dot\sigma+V_\sigma&=0,
    \label{eq:sigma-eom}\\
    \omega
    \equiv e_{sI}\dot e_\sigma^I
    &=-\frac{V_s}{\dot\sigma},
    \label{eq:turn-rate}
\end{align}
where the signed turn rate $\omega$ measures the bending of the trajectory.

We consider the benchmark parameters
\begin{equation}
 r_\Lambda=0.170,\quad \gamma=0.0072,\quad n_1=70\,M_{\rm Pl}^{-2}, \quad n_2=773.604\,M_{\rm Pl}^{-2}, \quad (\mathrm{BP\text{--}1})
 \label{eq:benchmark-BP1}
\end{equation} 
with $\Lambda$ fixed by the CMB normalization below.

For the numerical evolution, we use the e-fold variable $N$, defined by $dN=H\,dt$. All time-evolution panels use $N-N_{\rm t}$, where $N_{\rm t}$ denotes the end of the first inflationary stage, defined by the first upward crossing of $\epsilon_H=1$.

\begin{figure}[!tbp]
\centering
    \includegraphics[width=\linewidth]{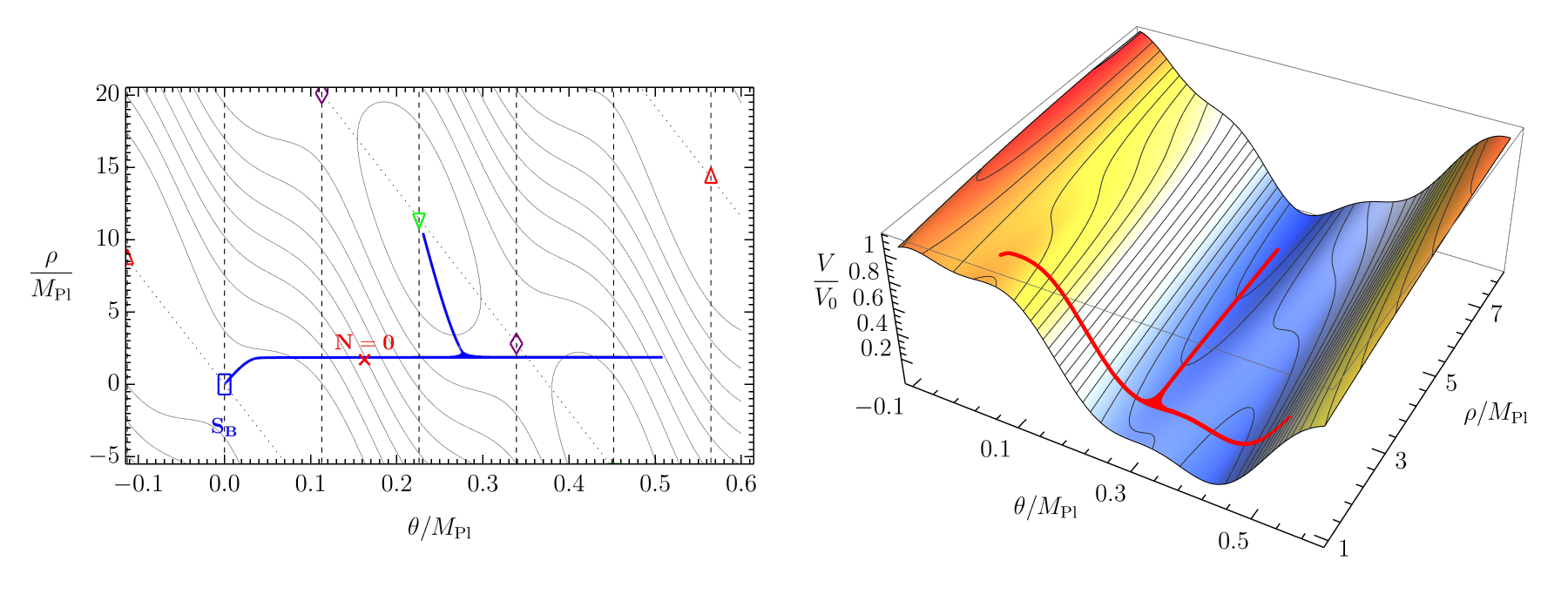}
\caption{
\textit{Left}: Potential contours and background trajectory for BP--1. The figure marks maxima (upward-pointing red triangles), minima (downward-pointing green triangles), saddle points $\mathcal S_A$ (purple diamonds), saddle points $\mathcal S_B$ (blue squares), and symmetry domains (dotted lines). \textit{Right}: The trajectory overlaid on the normalized potential, where $V_0$ is its value at the saddle. The first inflationary stage follows the nearly flat metastable branch. Loss of transverse stability redirects the fields into an adjacent valley, where the second inflationary stage takes place.}
\label{fig:trajectory_3d}
\end{figure}
\begin{figure}[!tbp]
\centering
    \includegraphics[width=\linewidth]{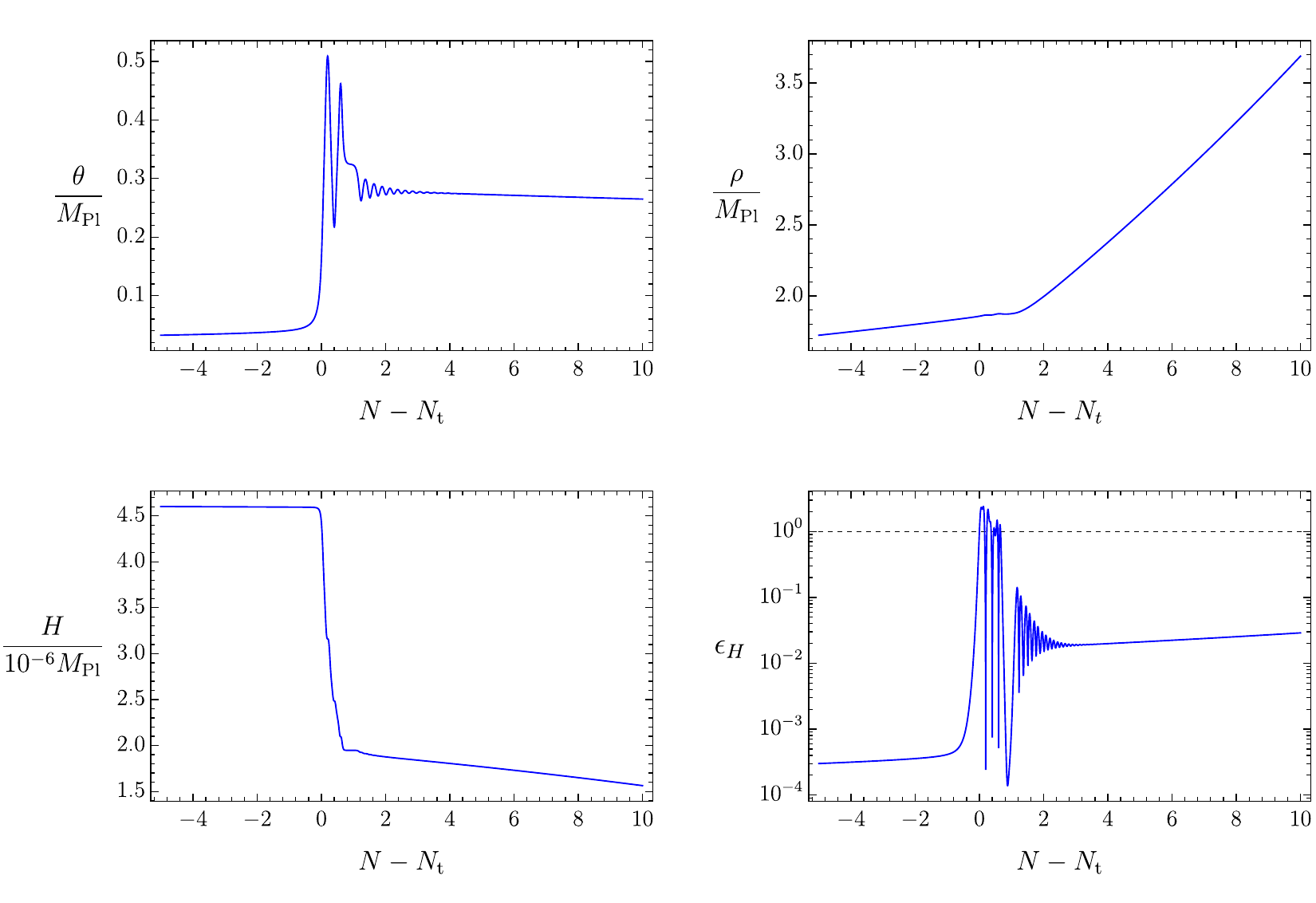}
\caption{
Background evolution for BP--1 as a function of $N-N_{\rm t}$, where $N_{\rm t}$ denotes the end of the first inflationary stage. Upper panels: the homogeneous fields $\theta$ and $\rho$. Lower panels: the Hubble parameter $H$ and $\epsilon_H$. The dashed line marks $\epsilon_H=1$; excursions above this threshold correspond to brief interruptions of inflation during the transition to the second stage.}
\label{fig:bckev}
\end{figure}

As shown in Fig.~\ref{fig:trajectory_3d}, the background initially follows the metastable branch near ${\cal S}_B$. The loss of transverse stability deflects the trajectory toward a neighbouring valley. The evolution is approximately straight during each inflationary stage, while the intermediate transition contains sharp turns.

Figure~\ref{fig:bckev} shows the rapid displacement and damped oscillations of $\theta$ as the fields settle into the second valley, where $\rho$ supports slow roll. The accompanying increase in kinetic energy briefly drives $\epsilon_H$ above unity, interrupting inflation. Inflation resumes as the kinetic energy is diluted. Such an intermediate noninflationary interval is familiar from double inflation~\cite{
Polarski:1992dq,Polarski:1994rz}. The intervals with $\epsilon_H>1$ span a total $\Delta N\simeq0.41$ during the transient relaxation.

To estimate the CMB observables at leading order in slow roll, we use
\begin{equation}
    \epsilon_V
    \equiv
    \frac{\mpl^2}{2}\frac{V_{,I}V_{,I}}{V^2},
    \qquad
    \eta_{\sigma\sigma}
    \equiv
    \mpl^2\frac{V_{\sigma\sigma}}{V}.
    \label{eq:potential-slow-roll}
\end{equation}
For BP--1, the CMB pivot exits during the approximately straight, slow-roll portion of the first stage, well before the instability. Neglecting subsequent entropy transfer on CMB scales, the leading predictions are
\begin{align}
    n_{s*}
    &\simeq
    1-6\epsilon_{V*}+2\eta_{\sigma\sigma *},
    \label{eq:slow-roll-scalar-tilt}\\
    r_*
    &\simeq16\epsilon_{V*},
    \label{eq:slow-roll-tensor-ratio}
\end{align}
where a star denotes evaluation at $k_*=a_*H_*=0.05\,{\rm Mpc}^{-1}$. We fix $\Lambda$ by requiring~\cite{Planck:2018jri}
\begin{equation}
    A_s
    \simeq
    \frac{V_*}{24\pi^2\mpl^4\epsilon_{V*}}
    =2.10\times10^{-9}.
    \label{eq:slow-roll-scalar-amplitude}
\end{equation}

Assuming instantaneous reheating followed by standard adiabatic thermal evolution, matching the pivot scale to the subsequent expansion history gives~\cite{Liddle:1993fq,Planck:2018jri}
\begin{equation}
    \Delta N_*
    \equiv
    \ln\frac{a_{\rm end}}{a_*}
    \simeq
    56.8+\frac14\ln\left(
        \frac{r_*V_*}{\varrho_{\rm end}}
    \right),
    \label{eq:pivot-matching}
\end{equation}
where $\varrho_{\rm end}=3\mpl^2H_{\rm end}^2$ is the total energy density at the final end of inflation. The numerical constant assumes $k_*/a_0=0.05\,{\rm Mpc}^{-1}$, $h=0.67$, $g_{*,\rm reh}=g_{*s,\rm reh}=106.75$, and the scalar normalization in Eq.~\eqref{eq:slow-roll-scalar-amplitude}. For both benchmarks considered in this work, $\Delta N_*\simeq57$.

For BP--1, we obtain $r_*\simeq2\times10^{-3}$, below the BICEP/Keck 2018 upper bound $r<0.036$~\cite{BICEP:2021xfz} and the combined Planck PR4, BICEP/Keck 2018, BAO, and CMB-lensing bound $r<0.032$~\cite{Tristram:2021tvh}, both at $95\%$ confidence level. The corresponding spectral index, $n_{s*}\simeq0.971$, is consistent with the joint ACT DR6, Planck and DESI BAO analysis~\cite{AtacamaCosmologyTelescope:2025blo}, as well as CMB-SPA+DESI DR2~\cite{McDonough:2025lzo}.

\section{Perturbations}
\label{sec:perturbations}

We study linear perturbations about the two-field background using the multifield formalism of Refs.~\cite{
Sasaki:1995aw,Gordon:2000hv,Wands:2002bn,Langlois:2008mn, Peterson:2010np,Gong:2011uw,Kaiser:2012ak,Gong:2016qmq}. We decompose the fields into homogeneous backgrounds and inhomogeneous perturbations,
\begin{equation}
    \phi^I(t,\boldsymbol{x})
    =
    \bar{\phi}^I(t)+\delta\phi^I(t,\boldsymbol{x}), \qquad
    \phi^I=(\theta,\rho),
    \label{eq:field-splitting}
\end{equation}
where $\bar{\phi}^I$ denotes the homogeneous background. In the following, we suppress the overbar on background quantities.

We introduce the gauge-invariant Mukhanov--Sasaki variables~\cite{Mukhanov:1988jd,Sasaki:1986hm},
\begin{equation}
    Q^I \equiv \delta\phi^I+\frac{\dot{\phi}^I}{H}\psi,
    \label{eq:QI-def}
\end{equation}
where $\psi$ denotes the scalar curvature perturbation of the spatial metric. Projecting onto the kinematic basis introduced in Eq.~\eqref{eq:kinematic-basis}, we define the adiabatic and entropy perturbations,
\begin{equation}
    Q_\sigma\equiv e_{\sigma I}Q^I, \qquad Q_s\equiv e_{sI}Q^I.
    \label{eq:Qs-decomp}
\end{equation}
For each Fourier mode with comoving wavenumber $k$, these
perturbations obey~\cite{Gordon:2000hv,GrootNibbelink:2001qt}
\begin{equation}
    \ddot Q_\sigma +3H\dot Q_\sigma +\left(\frac{k^2}{a^2}+\mu_\sigma^2 \right)Q_\sigma = 2\frac{\dd}{\dd t}\left(\omega Q_s\right)-2\left(\frac{V_\sigma}{\dot\sigma}+\frac{\dot H}{H} \right)\omega Q_s,
    \label{eq:Qsigma-eom}
\end{equation}
and
\begin{equation}
    \ddot Q_s +3H\dot Q_s +\left(\frac{k^2}{a^2}+\mu_s^2 \right)Q_s = 4\mpl^2 \frac{\omega}{\dot\sigma} \frac{k^2}{a^2}\Psi_{\rm B},
    \label{eq:entropy-eom}
\end{equation}
where $\Psi_{\rm B}$ is the gauge-invariant Bardeen
potential. The diagonal mass terms are
\begin{align}
    \mu_\sigma^2 &\equiv V_{\sigma\sigma}-\omega^2-\frac{1}{\mpl^2a^3} \frac{\dd}{\dd t} \left(\frac{a^3\dot\sigma^2}{H} \right),
    \label{eq:effective-adiabatic-mass}\\
    \mu_s^2 &\equiv  V_{ss}+3\omega^2.
    \label{eq:effective-entropy-mass}
\end{align}
The comoving curvature and dimensionless entropy perturbations are
\begin{equation}
    \mathcal R \equiv \frac{H}{\dot\sigma}Q_\sigma,
    \qquad
    S \equiv \frac{H}{\dot\sigma}Q_s.
    \label{eq:R-S-def}
\end{equation}
For each wavenumber, we evolve two independent solutions initialized in the Bunch--Davies vacuum deep inside the Hubble radius and sum their contributions to the power spectra~\cite{Bunch:1978yq,Lalak:2007vi}. Their field-space components initially satisfy
\begin{equation}
    Q^I_\lambda(\tau,k) \longrightarrow \frac{e^{-ik\tau}}{a\sqrt{2k}}\, \delta^I_\lambda,
    \label{eq:BD-initial}
\end{equation}
where $\tau$ is conformal time and $\lambda=1,2$ labels the independent initial field fluctuations. Projecting the solutions onto the kinematic basis gives $\mathcal R_\lambda$ and $S_\lambda$, with power spectra
\begin{equation}
    \mathcal P_{\mathcal R}(k,N)
    =
    \frac{k^3}{2\pi^2}
    \sum_{\lambda=1}^{2}
    \left|\mathcal R_\lambda(k,N)\right|^2,
    \qquad
    \mathcal P_S(k,N)
    =
    \frac{k^3}{2\pi^2}
    \sum_{\lambda=1}^{2}
    \left|S_\lambda(k,N)\right|^2.
    \label{eq:spectra-def}
\end{equation}
We identify the primordial curvature spectrum with its value at the final end of inflation,
\begin{equation}
    \mathcal P_\zeta(k)
    \equiv
    \mathcal P_{\mathcal R}(k,N_{\rm f}),
    \label{eq:primordial-spectrum}
\end{equation}
where $N_{\rm f}$ denotes the end of the second inflationary stage, after the entropy source has become negligible and the curvature spectrum has approached a constant.

\begin{figure}[!tbp]
\centering
    \includegraphics[
        width=\linewidth,
        height=6cm,
        keepaspectratio
    ]{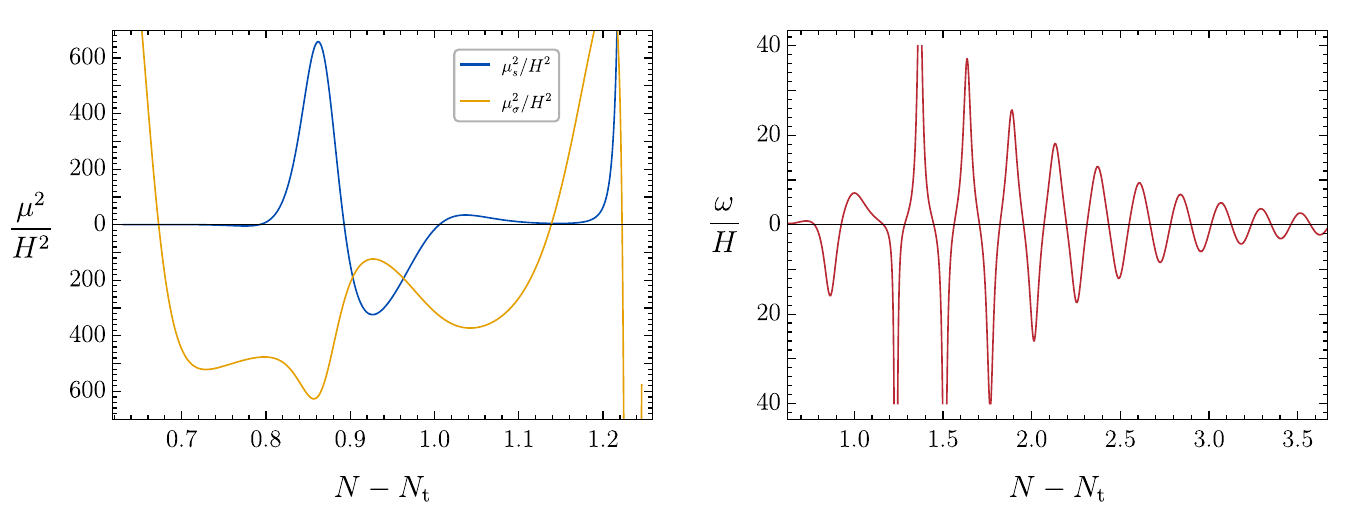}

\caption{Effective masses and turning for BP--1 as functions
of $N-N_{\rm t}$.  \textit{Left}: entropy mass $\mu_s^2/H^2$ (blue) and adiabatic mass $\mu_\sigma^2/H^2$ (orange). Both become negative during the transition, allowing growth when the negative mass terms overcome the gradient contributions. \textit{Right}: the turn rate $\omega/H$ along the benchmark trajectory, which controls the coupling between adiabatic and entropy perturbations. Its damped oscillations accompany the relaxation into the second inflationary valley.}
\label{fig:entropy-mass-turn-rate}
\end{figure}

The projected potential curvatures are
\begin{equation}
    V_{aa} = \frac{\Lambda^4}{1+r_\Lambda}
    \left[-\cos{\cal A}_1 (\boldsymbol q_1\cdot\boldsymbol e_a)^2 +r_\Lambda\cos{\cal A}_2 (\boldsymbol q_2\cdot\boldsymbol e_a)^2 \right], \qquad a=\sigma,s.
    \label{eq:Vss-origin}
\end{equation}
Near ${\cal S}_B$, both cosines are positive, so the first contribution is destabilizing and the second is restoring. Their balance depends on the direction onto which the Hessian is projected. The trajectory initially follows the light tachyonic direction identified in Sec.~\ref{sec:model}, giving a small negative $V_{\sigma\sigma}$, while the second contribution dominates in the transverse direction and maintains $V_{ss}>0$.

Along the metastable branch, the transverse restoring contribution weakens as the phases evolve. When it no longer compensates the destabilizing term, $V_{ss}$ becomes negative and transverse stabilization is lost. During the subsequent transition, both the cosine factors and their projections change as the fields move across the potential and the trajectory turns. The instantaneous tangent can then acquire a larger projection onto negative-curvature directions, producing more pronounced negative excursions of $V_{\sigma\sigma}$. The effective perturbation masses also depend on turning: $\mu_s^2$ contains the positive contribution $3\,\omega^2$, whereas $\mu_\sigma^2$ contains $-\omega^2$ and the gravitational correction in Eq.~\eqref{eq:effective-adiabatic-mass}.

Figure~\ref{fig:entropy-mass-turn-rate} shows the effective entropy and adiabatic masses (left panel) and the turn rate $\omega/H$ (right panel) for BP--1. The adiabatic mass becomes substantially negative before the deepest negative excursion of the entropy mass. These intervals can support growth of $Q_\sigma$ and $Q_s$ when the corresponding negative mass terms overcome the gradient contribution $k^2/a^2$. At the same time, turning couples the two perturbations, so their amplification cannot be inferred from the diagonal masses alone. The subsequent damped oscillations of $\omega/H$ reflect repeated changes in the direction of motion as the fields relax into the second inflationary valley. We follow this evolution using the full coupled system.

To illustrate the dependence of the small-scale enhancement on the model parameters, we consider a second benchmark solution, BP--2. It has the same values of $n_1$ and $\gamma$ as BP--1 in Eq.~\eqref{eq:benchmark-BP1}, but differs in $n_2$ and
$r_\Lambda$,
\begin{equation}
    r_\Lambda=0.190,
    \qquad
    n_2=744.066\,M_{\rm Pl}^{-2}
    \qquad
    (\mathrm{BP\text{--}2}) .
    \label{eq:benchmark-BP2}
\end{equation}
For this benchmark, the overall scale $\Lambda$ is again fixed by the CMB normalization in Eq.~\eqref{eq:slow-roll-scalar-amplitude}.

\begin{figure}[t]
\centering
    \includegraphics[
        width=\linewidth,
        height=6cm,
        keepaspectratio
    ]{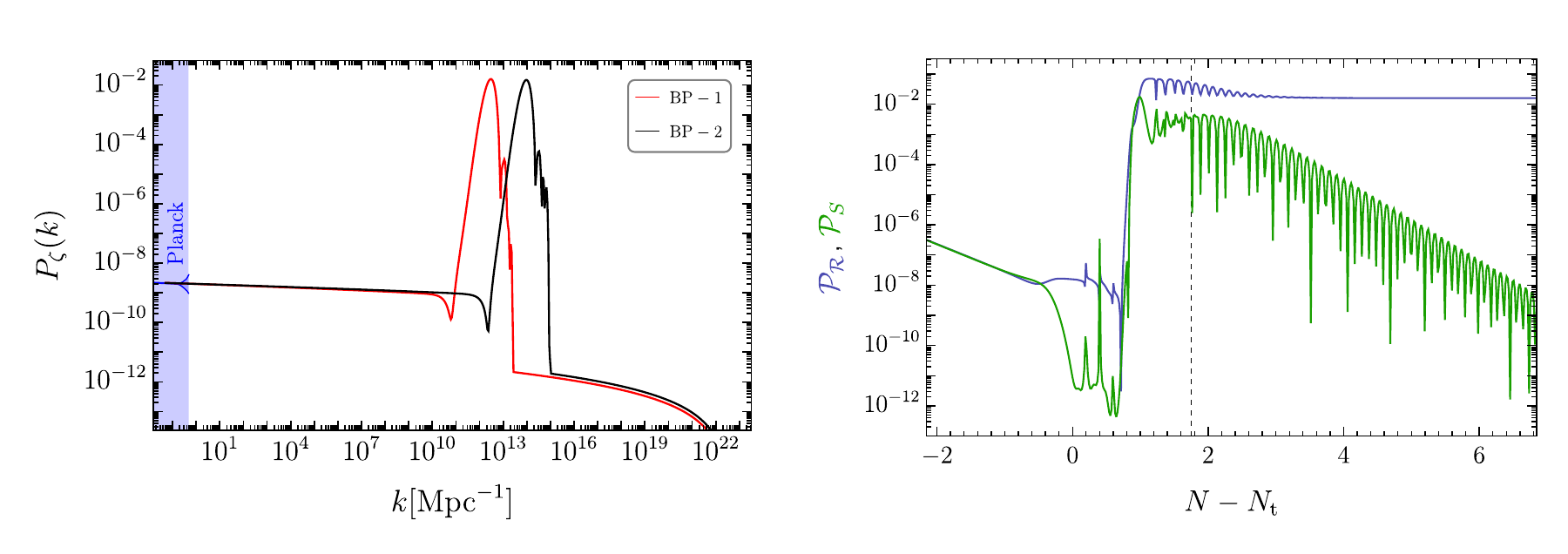}
\caption{\textit{Left}: final primordial curvature spectra $\Pzeta(k)$ for BP--1 (red) and BP--2 (black), showing localized small-scale enhancements. \textit{Right}: curvature power $\mathcal P_{\mathcal R}$ (blue) and entropy power $\mathcal P_S$ (green) for the BP--1 peak mode as functions of $N-N_{\rm t}$. Both are amplified during the transition; subsequently, the entropy power decays while the curvature power approaches a constant. The vertical dashed line marks Hubble exit, $k=aH$, for this mode.}
\label{fig:ps}
\end{figure}

The left panel of Fig.~\ref{fig:ps} shows the final primordial curvature spectra for BP--1 and BP--2. Both benchmarks preserve the nearly scale invariant spectrum on large scales while producing a localized enhancement of order $10^{-2}$ on smaller scales. The numerical curvature power spectra obtained with our code were cross-checked against the publicly available Mathematica package \texttt{mTransport}~\cite{Dias:2015rca}, which implements the transport formalism for multifield inflation~\cite{Seery:2012vj}, with consistent results. The right panel shows the evolution of the curvature and entropy power spectra for the mode located near the peak of BP--1. During the transition, the entropy power (green) grows rapidly, while turning couples the entropy and adiabatic perturbations and the curvature power (blue) is amplified. After the transition, the entropy mode undergoes damped oscillations and eventually becomes negligible, whereas the curvature power approaches a conserved plateau. For BP--2, the second inflationary stage is shorter and the localized enhancement is shifted towards larger $k$, while its peak amplitude remains comparable to that of BP--1.

To assess the effect of entropy sourcing, we evolve an auxiliary adiabatic solution with the right-hand side of Eq.~\eqref{eq:Qsigma-eom} set to zero, retaining the same background, diagonal mass and initial adiabatic data. For the BP--1 peak mode, this gives $\mathcal P_\zeta^{(0)}\simeq6.71\times10^{-3}$, compared with $1.59\times10^{-2}$ in the full system. Including entropy sourcing therefore increases the final power by a factor of approximately $2.37$. For BP--2, the corresponding values are $\mathcal P_\zeta^{(0)}\simeq2.54\times10^2$ and $\mathcal P_\zeta\simeq1.49\times10^{-2}$: the source-induced solution strongly interferes destructively with the auxiliary solution. Destructive multifield effects have been discussed, for example, in Ref.~\cite{McAllister:2012am}. Here, interference occurs between the auxiliary and source-induced parts of each mode solution, whereas the powers of independent initial vacuum modes add without interference. The auxiliary evolution retains the diagonal effects of turning while omitting the associated entropy coupling. This diagnostic therefore demonstrates the importance of mixing, rather than isolating a physical contribution attributable to negative $\mu_\sigma^2$.

For the modes at the peak of the curvature spectrum, we find $k_{\rm peak}/(aH)\simeq2.31$ and $2.30$ for BP--1 and BP--2, respectively, evaluated at the deepest minimum of the effective entropy mass during the transition. The amplification thus involves modes that are mildly inside the Hubble radius during this interval. This is relevant to the nonlinear rescattering discussed in Refs.~\cite{Fumagalli:2023loc,Inomata:2022yte,Braglia:2026fzp}, whose resonant enhancement is particularly pronounced when the fluctuations are excited far inside the Hubble radius. The corresponding deeply sub-Hubble approximation is therefore not directly applicable to this stage of our benchmarks. A quantitative assessment of the nonlinear corrections in the present model requires a dedicated calculation beyond the scope of this work.

%

\section{Primordial-black-hole formation}
\label{sec:pbh}

The enhancement of the primordial curvature spectrum generated during the transition between the two inflationary stages can seed the formation of primordial black holes when the corresponding modes re-enter the Hubble horizon during radiation domination~\cite{Hawking:1971ei,Carr:1974nx, Escriva:2021aeh,Escriva:2022duf}. Using the spectrum defined in Eq.~\eqref{eq:primordial-spectrum}, we compute the PBH abundance assuming Gaussian primordial curvature perturbations, while retaining the nonlinear relation between curvature and compaction. Primordial non-Gaussianity generated by interactions during the transition may also affect the tail of the distribution relevant for collapse~\cite{Franciolini:2018vbk,Ferrante:2022mui}. A detailed investigation of these effects lies beyond the scope of the present work.

We evaluate the collapse probability using the compaction function~\cite{Shibata:1999zs}, following the prescription summarized in Appendix~\ref{app:pbh_details}. During radiation domination, the nonlinear compaction function at its maximum can be expressed in terms of its linear component $\mathcal C_1$ as~\cite{Harada:2015yda,Young:2019yug}
\begin{equation}
    \mathcal C
    =
    \mathcal C_1
    -\frac{\mathcal C_1^{\,2}}{4\Phi},
    \label{eq:nonlinear_compaction}
\end{equation}
with $\Phi=2/3$ during radiation domination. A perturbation collapses when $\mathcal C>\mathcal C_{\rm th}$. The threshold $\mathcal C_{\rm th}$ depends on the shape of the curvature spectrum~\cite{Germani:2018jgr,Musco:2018rwt,Musco:2020jjb} and is determined separately for each benchmark using the procedure described in Appendix~\ref{app:pbh_threshold}.

Close to the threshold, the PBH mass obeys the critical-scaling relation~\cite{Choptuik:1992jv,Evans:1994pj,Musco:2008hv,Niemeyer:1997mt}
\begin{equation}
    M
    =
    \mathcal K M_H
    \left(\mathcal C-\mathcal C_{\rm th}\right)^{\nu_c},   \label{eq:critical_scaling}
\end{equation}
where $M_H$ is the horizon mass at formation. We adopt $\mathcal K\simeq4$ and $\nu_c\simeq0.36$, appropriate to collapse during radiation domination~\cite{Musco:2008hv}. The characteristic mass associated with a mode $k$ may be written as
\begin{align}
    M_{\rm PBH}(k)
    \simeq
    1.7\times10^{-12}M_\odot\,
    \left(\frac{\gamma_{\rm coll}}{0.6}\right)
    \left(\frac{106.75}{g_*(T_E)}\right)^{1/6}
    \left[
        \frac{2.24\times10^{12}\ {\rm Mpc}^{-1}}
        {k/(\kappa a_0)}
    \right]^2 ,
    \label{eq:k_to_pbh_mass}
\end{align}
where $\gamma_{\rm coll}\equiv M_{\rm PBH}/M_H$ denotes the effective mean collapse efficiency, $r_m=\kappa/k$ is the characteristic scale of the perturbation, and $\kappa=\mathcal O(3)$~\cite{Musco:2018rwt,Germani:2018jgr,Escriva:2019phb,Young:2019osy}. For a sufficiently narrow spectrum with Gaussian statistics, $\gamma_{\rm coll}\simeq0.6$ provides a representative value.

The variance of the linear compaction variable $\mathcal C_1$ on the scale $r_m$ is~\cite{Young:2019yug}
\begin{align}
    \sigma_{\mathcal C}^{2}(r_m)
    =
    \left(\frac{2\Phi}{3}\right)^2
    \int_0^\infty \frac{dq}{q}\,
    (q r_m)^4\,
    W^2(qr_m)\,
    T^2(q,\eta_m)\,
    \mathcal P_{\zeta}(q),
    \label{eq:compaction_variance}
\end{align}
where $\eta_m=r_m$ denotes the conformal time
at which the characteristic scale enters the Hubble radius
during radiation domination and 
\begin{align}
    W(x)
    &=
    3\,\frac{\sin x-x\cos x}{x^3},
    \label{eq:top-hat-window}
    \\
    T(q,\eta)
    &=
    3\,
    \frac{
        \sin(q\eta/\sqrt{3})
        -(q\eta/\sqrt{3})\cos(q\eta/\sqrt{3})
    }{
        (q\eta/\sqrt{3})^3
    },
    \label{eq:radiation-transfer}
\end{align}
are the Fourier transform of the real-space top-hat window and the radiation-era transfer function, respectively. Under the Gaussian assumption,
\begin{equation}
    P(\mathcal C_1;r_m)
    =
    \frac{1}{\sqrt{2\pi}\sigma_{\mathcal C}(r_m)}
    \exp\left[
        -\frac{\mathcal C_1^{\,2}}
        {2\sigma_{\mathcal C}^{2}(r_m)}
    \right].
    \label{eq:Gaussian_compaction_pdf}
\end{equation}

The collapse threshold expressed in terms of the Gaussian compaction variable follows from $\mathcal C(\mathcal C_{1,{\rm th}})=\mathcal C_{\rm th}$. On the type-I branch, it is given by
\begin{equation}
    \mathcal C_{1,{\rm th}}
    =
    2\Phi
    \left(
        1-\sqrt{1-\frac{\mathcal C_{\rm th}}{\Phi}}
    \right).
    \label{eq:Gaussian_compaction_threshold}
\end{equation}
The fraction of the total energy density collapsing into PBHs at formation, for a fixed horizon mass $M_H$, is~\cite{Young:2019yug}
\begin{align}
    \beta(M_H)
    &=
    \int_{\mathcal C_{1,{\rm th}}}^{2\Phi}
    \mathrm d\mathcal C_1\,
    \frac{M(\mathcal C_1,M_H)}{M_H}\,
    P(\mathcal C_1;r_m)
    \nonumber\\
    &=
    \mathcal K
    \int_{\mathcal C_{1,{\rm th}}}^{2\Phi}
    \mathrm d\mathcal C_1\,
    \left[
        \mathcal C_1
        -\frac{\mathcal C_1^{\,2}}{4\Phi}
        -\mathcal C_{\rm th}
    \right]^{\nu_c}
    P(\mathcal C_1;r_m).
    \label{eq:beta_gaussian}
\end{align}
The lower limit enforces the collapse condition, while the upper limit restricts the integration to the type-I branch of the nonlinear compaction relation. In the second equality, the critical-scaling relation~\eqref{eq:critical_scaling} has been used.

The present-day differential PBH fraction is defined as
\begin{equation}
    f_{\rm PBH}(M)
    \equiv
    \frac{1}{\rho_{{\rm DM},0}}
    \frac{\dd\rho_{{\rm PBH},0}}{\dd\ln M},
    \label{eq:fpbh_definition}
\end{equation}
and its integral gives the total fraction of dark matter in PBHs,
\begin{align}
    f_{\rm PBH}^{\rm tot}
    &\equiv
    \int \dd\ln M\,f_{\rm PBH}(M)
    \nonumber\\
    &=
    \frac{1}{\Omega_{\rm DM}}
    \int_{M_{H,\min}}^{M_{H,\max}}
    \dd\ln M_H\,
    \left(\frac{M_{\rm eq}}{M_H}\right)^{1/2}
    \beta(M_H).
    \label{eq:total_fpbh}
\end{align}
Here $M_{\rm eq}$ denotes the horizon mass at matter--radiation equality. For the corresponding differential PBH mass function, see Appendix~\ref{app:pbh_mass_function}.

Figure~\ref{fig:pbh-abundance} shows the PBH mass functions obtained from the benchmark curvature spectra. Their location is set primarily by the wavenumber of the scalar peak through $M\propto k^{-2}$, while their width is generated by both the finite width of $\Pzeta(k)$ and critical collapse. For the benchmarks shown, the distributions fall in the asteroid-mass window in which the displayed observational limits permit an appreciable, and potentially dominant, PBH dark-matter component.

\begin{figure}[t]
    \centering
        \includegraphics[
            width=0.7\linewidth,
        ]{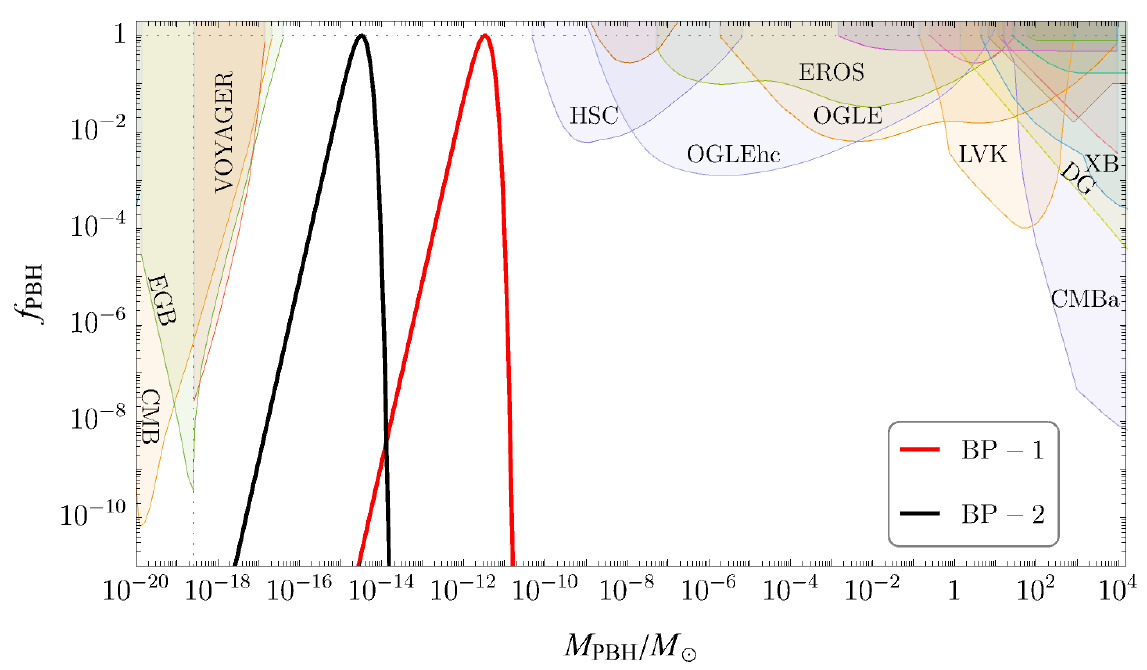}
    
    \caption{PBH mass functions for BP--1 (red) and BP--2 (black). For the benchmarks shown, the mass functions lie in the asteroid-mass region and are displayed together with the monochromatic constraints for reference compiled in Ref.~\cite{Carr:2026hot}.}
 \label{fig:pbh-abundance}
\end{figure}
%

\section{Scalar-induced gravitational waves}
\label{sec:sigw}

The enhancement of the primordial curvature spectrum also generates a stochastic gravitational-wave background through the second-order coupling between scalar and tensor perturbations~\cite{Matarrese:1992rp,Matarrese:1997ay,Bruni:1996im,Carbone:2004iv,Ananda:2006af,Baumann:2007zm}. Decomposing the transverse-traceless tensor perturbation into its two polarizations, $\lambda=+,\times$, each Fourier mode obeys
\begin{equation}
    h_\lambda''(\eta,\mathbf{k})
    +2\mathcal H h_\lambda'(\eta,\mathbf{k})
    +k^2h_\lambda(\eta,\mathbf{k})
    =
    4\mathcal S_\lambda(\eta,\mathbf{k}),
    \label{eq:sourced-tensor-eom}
\end{equation}
where primes denote derivatives with respect to conformal time, $\mathcal H\equiv a'/a$, and $\mathcal S_\lambda$ is the second-order source constructed from products of first-order scalar perturbations.

For modes re-entering during radiation domination, we define the late-time, oscillation-averaged tensor quantity
\begin{equation}
    \widehat{\mathcal P}_h(k)
    \equiv
    \lim_{x\to\infty}
    x^2\,\overline{\mathcal P_h(k,\eta)},
    \qquad x\equiv k\eta,
    \label{eq:late-time-tensor-definition}
\end{equation}
where the overbar denotes an average over tensor oscillations. The factor $x^2$ removes the radiation-era decay of the tensor power spectrum.
Assuming Gaussian primordial curvature perturbations, this quantity is given by~\cite{Espinosa:2018eve,Kohri:2018awv}
\begin{equation}
    \widehat{\mathcal P}_h(k)
    =
    4
    \int_1^\infty \dd t
    \int_0^1 \dd s\,
    \left[
        \frac{(t^2-1)(1-s^2)}{t^2-s^2}
    \right]^2
    {\cal I}_{t,s}^2\,
    \Pzeta\!\left(\frac{k}{2}(t-s)\right)
    \Pzeta\!\left(\frac{k}{2}(t+s)\right),
    \label{eq:P_h_ts}
\end{equation}
where $t=u+v$ and $s=u-v$, with
\begin{equation}
    v\equiv\frac{|\mathbf p|}{k},
    \qquad
    u\equiv\frac{|\mathbf{k}-\mathbf{p}|}{k}.
    \label{eq:sigw-integration-variables}
\end{equation}
The radiation-era kernel, including the late-time rescaling and oscillation average, is
\begin{align}
    {\cal I}_{t,s}^2
    ={}&
    \frac{288(s^2+t^2-6)^2}{(t^2-s^2)^6}
    \Bigg\{
        \frac{\pi^2}{4}
        (s^2+t^2-6)^2
        \Theta(t-\sqrt{3})
    \nonumber\\
    &\hspace{2.5cm}
        +
        \left[
            t^2-s^2
            -\frac{s^2+t^2-6}{2}
            \ln\left|
                \frac{t^2-3}{3-s^2}
            \right|
        \right]^2
    \Bigg\}.
    \label{eq:I_RD_osc_ave_ts}
\end{align}

\begin{figure}[t]
    \centering
    \includegraphics[
        width=0.7\linewidth,
    ]{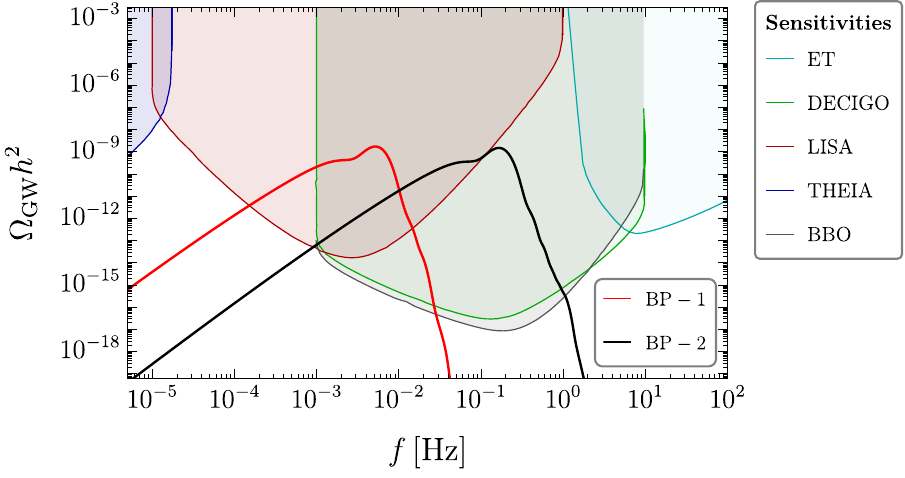}
    \caption{Present-day SIGW energy-density spectra for BP--1 (red) and BP--2 (black). The projected sensitivity curves of ET, DECIGO, LISA, THEIA, and BBO are shown for comparison.}
    \label{fig:sigw-spectrum}
\end{figure}

The corresponding present-day gravitational-wave energy density is
\begin{equation}
    h^2\Omega_{{\rm GW},0}(k)
    =
    \frac{h^2\Omega_{r,0}}{24}
    \frac{g_*(T_k)}{g_{*,0}}
    \left[
        \frac{g_{*s}(T_k)}{g_{*s,0}}
    \right]^{-4/3}
    \widehat{\mathcal P}_h(k),
    \label{eq:GW}
\end{equation}
where $T_k$ denotes the temperature at horizon re-entry, while $g_*$ and $g_{*s}$ are the effective relativistic degrees of freedom associated with the energy and entropy densities, respectively. We use $h^2\Omega_{r,0}=4.2\times10^{-5}$.

As shown in Fig.~\ref{fig:sigw-spectrum}, the BP--1 signal peaks in the millihertz band, overlapping the projected sensitivity of LISA~\cite{LISA:2017pwj}. Dedicated LISA forecasts for inflationary signals and the reconstruction of their scalar sources have been developed in Refs.~\cite{LISACosmologyWorkingGroup:2024hsc,LISACosmologyWorkingGroup:2025vdz}. BP--2 instead peaks in the decihertz band, overlapping the projected sensitivities of DECIGO~\cite{Kawamura:2020pcg} and BBO~\cite{Corbin:2005ny}. These sensitivity comparisons are indicative; we do not perform a likelihood analysis including detector noise and astrophysical foregrounds.

\section{Conclusions}
\label{sec:conclusion}

We have investigated the small-scale consequences of inflationary trajectories that begin in the metastable regime of aligned axion inflation. During the first slow-roll stage, the backgrounds follow a CMB-compatible trajectory near a saddle point. When transverse stability is lost, the fields undergo a short nonadiabatic transition before settling into a neighbouring valley and entering a second period of inflation.

By numerically solving the complete coupled system of adiabatic and entropy modes, we find that negative effective masses and strong turning accompany a localized enhancement of the primordial curvature spectrum, while its nearly scale-invariant form is preserved on CMB scales. The amplification depends on the coupled adiabatic--entropy evolution through the transition. The two benchmark solutions further show that the characteristic scale of the enhancement can be shifted by varying the duration of the second inflationary stage while maintaining a comparable peak amplitude.

We have used the numerical curvature spectra to compute the corresponding PBH mass functions assuming Gaussian primordial curvature perturbations, within the nonlinear compaction-function formalism, determining the collapse threshold from the shape of each spectrum and including critical collapse. The benchmark solutions produce extended PBH distributions in the asteroid-mass range, with a potentially significant dark-matter abundance. We further computed the corresponding present-day SIGW spectra for both benchmarks. The PBH mass function and gravitational-wave spectrum therefore provide complementary probes of the perturbations generated during the transition. These results show that the phenomenology of the minimal aligned axion model extends beyond its predictions for $n_s$ and $r$.

\acknowledgments

The author thanks Matteo Braglia and Marco Peloso for carefully reading the draft and for their valuable comments. The author acknowledges support from the Istituto Nazionale di Fisica Nucleare (INFN) through the Theoretical Astroparticle Physics (TAsP) project.


\appendix

\section{Details of the primordial-black-hole calculation}
\label{app:pbh_details}

This appendix collects the supplementary relations used to associate the primordial curvature spectrum with the PBH mass and abundance.

\subsection{Mass--scale relation}
\label{app:pbh_mass_scale}

During radiation domination, the energy density of the thermal bath at horizon entry is $\rho_E=(\pi^2/30)g_{*,E}T_E^4$, where the subscript $E$ denotes quantities evaluated at the horizon-entry time of the scale $r_m$. Combining this relation with the Friedmann equation $H_E^2=\rho_E/(3M_{\rm Pl}^2)$ gives
\begin{equation}
    H_E
    =
    \sqrt{\frac{\pi^2g_{*,E}}{90}}\,
    \frac{T_E^2}{M_{\rm Pl}}.
    \label{eq:hubble_entry_appendix}
\end{equation}
The mass contained within a Hubble sphere at horizon entry is therefore
\begin{equation}
    M_H
    =
    \frac{4\pi}{3}\rho_E H_E^{-3}
    =
    4\pi\frac{M_{\rm Pl}^2}{H_E}
    =
    \frac{12\sqrt{10}\,M_{\rm Pl}^{3}}
    {g_{*,E}^{1/2}T_E^{2}}.
    \label{eq:horizon_mass_appendix}
\end{equation}
Conservation of comoving entropy implies $g_{*s}^{1/3}aT=g_{*s,0}^{1/3}a_0T_0$, while the characteristic scale $r_m$ enters the horizon when $a_EH_Er_m=1$. Using $r_m=\kappa/k$, these relations give the temperature at horizon entry as
\begin{equation}
    T_E
    =
    \frac{3\sqrt{10}}{\pi}
    \frac{1}{g_{*,E}^{1/2}}
    \left(
        \frac{g_{*s,E}}{g_{*s,0}}
    \right)^{1/3}
    \frac{M_{\rm Pl}}{T_0}
    \frac{k}{\kappa a_0}.
    \label{eq:entry_temperature}
\end{equation}
For modes entering before neutrino decoupling, one may set $g_{*s,E}=g_{*,E}$. Equation~\eqref{eq:entry_temperature} then becomes
\begin{equation}
    T_E
    \simeq
    130\,{\rm TeV}\,
    \left(
        \frac{106.75}{g_{*,E}}
    \right)^{1/6}
    \left[
        \frac{k/(\kappa a_0)}
        {2.24\times10^{12}\,{\rm Mpc}^{-1}}
    \right].
    \label{eq:entry_temperature_numerical}
\end{equation}
Substituting Eq.~\eqref{eq:entry_temperature_numerical} into Eq.~\eqref{eq:horizon_mass_appendix} and multiplying by the mean collapse efficiency $\gamma_{\rm coll}$ gives Eq.~\eqref{eq:k_to_pbh_mass}, with $M_{\rm PBH}\propto k^{-2}$.
\subsection{Collapse threshold}
\label{app:pbh_threshold}

For a spherically symmetric perturbation, the nonlinear density contrast in the long-wavelength approximation is given by~\cite{Harada:2015yda}
\begin{equation}
    \frac{\delta\rho}{\rho_b}
    =
    -\frac{2\Phi}{3}
    \frac{1}{(aH)^2}
    e^{-2\zeta(r)}
    \left[
        \zeta''(r)
        +\frac{2}{r}\zeta'(r)
        +\frac{1}{2}\zeta'^{\,2}(r)
    \right].
    \label{eq:nonlinear_density_appendix}
\end{equation}
The compaction function is defined as twice the mass excess within the areal radius $R(r)$, divided by that radius,
\begin{equation}
    \mathcal C(r)
    \equiv
    \frac{2G\,[M(r)-M_b(r)]}{R(r)}.
\end{equation}
At its maximum $r=r_m$, it is given by Eq.~\eqref{eq:nonlinear_compaction} with
\begin{equation}
    \mathcal C_1
    =
    -2\Phi r_m\zeta'(r_m).
\end{equation}
On the type-I branch, $0\leq\mathcal C_1\leq2\Phi$, the nonlinear compaction is monotonic and reaches its maximum value $\mathcal C_{\max}=\Phi$ at $\mathcal C_1=2\Phi$.

For each benchmark, the collapse threshold is obtained from its numerical curvature spectrum through the following steps~\cite{Musco:2020jjb}:
\begin{enumerate}
    \item We first define
    \begin{equation}
        P_\zeta(k,\eta)
        \equiv
        \frac{2\pi^2}{k^3}
        \Pzeta(k)T^2(k,\eta),
    \end{equation}
    where $T(k,\eta)$ is the radiation-era transfer function defined in Eq.~\eqref{eq:radiation-transfer}. We evaluate it at $\eta=r_m/10$, for which the characteristic scale $r_m$ remains well outside the comoving Hubble radius. The characteristic scale $r_m$ is determined by
    \begin{align}
        0={}&
        \int_0^\infty \dd k\,k^2
        \left[
            (k^2r_m^2-1)
            \frac{\sin(kr_m)}{kr_m}
            +\cos(kr_m)
        \right]
         P_\zeta(k,\eta).
        \label{eq:rm_condition}
    \end{align}

    \item The Gaussian shape parameter is evaluated as
    \begin{align}
        \alpha_{\rm G}
        =
        -\frac{1}{4}
        \left[
            1+
            r_m
            \frac{
                \displaystyle
                \int_0^\infty \dd k\,k^4
                \cos(kr_m) P_\zeta(k,\eta)
            }{
                \displaystyle
                \int_0^\infty \dd k\,k^3
                \sin(kr_m) P_\zeta(k,\eta)
            }
        \right].
        \label{eq:alpha_gaussian}
    \end{align}

    \item The corresponding nonlinear shape parameter $\alpha$ is obtained by solving
    \begin{equation}
        F(\alpha)
        \left[1+F(\alpha)\right]
        \alpha
        =
        2\alpha_{\rm G},
        \label{eq:alpha_nonlinear}
    \end{equation}
    where
    \begin{equation}
        F(\alpha)
        =
        \left[
            1-\frac{2}{5}
            e^{-1/\alpha}
            \frac{
                \alpha^{\,1-5/(2\alpha)}
            }{
                \Gamma\!\left(\frac{5}{2\alpha}\right)
                -
                \Gamma\!\left(
                    \frac{5}{2\alpha},
                    \frac{1}{\alpha}
                \right)
            }
        \right]^{1/2}.
    \end{equation}

    \item Finally, the collapse threshold is approximated by
    \begin{equation}
        \mathcal C_{\rm th}(\alpha)
        \simeq
        \frac{4}{15}
        \frac{
            e^{-1/\alpha}
            \alpha^{\,1-5/(2\alpha)}
        }{
            \displaystyle
            \Gamma\!\left(\frac{5}{2\alpha}\right)
            -
            \Gamma\!\left(
                \frac{5}{2\alpha},
                \frac{1}{\alpha}
            \right)
        }.
        \label{eq:shape_dependent_threshold}
    \end{equation}
\end{enumerate}
Applying this procedure separately to the numerical curvature spectrum of each benchmark yields nearly identical thresholds, $\mathcal C_{\rm th}\simeq0.573$.

\subsection{PBH mass function}
\label{app:pbh_mass_function}

Combining the critical-scaling relation \eqref{eq:critical_scaling} with the nonlinear compaction relation \eqref{eq:nonlinear_compaction}, the linear compaction variable can be expressed in terms of $M$ and $M_H$. On the type-I branch, the solution is
\begin{equation}
    \mathcal C_1(M,M_H)
    =
    2\Phi
    \left[
        1-
        \sqrt{
            1-\frac{\mathcal C_{\rm th}}{\Phi}
            -\frac{1}{\Phi}
            \left(
                \frac{M}{\mathcal K M_H}
            \right)^{1/\nu_c}
        }
    \right].
    \label{eq:C1_inversion}
\end{equation}
The differential present-day mass function is therefore
\begin{align}
    f_{\rm PBH}(M)
    ={}&
    \frac{1}{\Omega_{\rm DM}}
    \int_{M_H^{\rm min}(M)}^\infty d\ln M_H\,
    \left(\frac{M_{\rm eq}}{M_H}\right)^{1/2}
    \left[
        1-\frac{\mathcal C_{\rm th}}{\Phi}
        -\frac{1}{\Phi}
        \left(
            \frac{M}{\mathcal K M_H}
        \right)^{1/\nu_c}
    \right]^{-1/2}
    \nonumber\\
    &\times
    \frac{\mathcal K}{\nu_c}
    \left(
        \frac{M}{\mathcal K M_H}
    \right)^{(1+\nu_c)/\nu_c}
    P\!\left(
        \mathcal C_1(M,M_H);r_m(M_H)
    \right).
    \label{eq:full_differential_mass_function}
\end{align}
The lower integration limit follows from requiring the radicand in Eq.~\eqref{eq:C1_inversion} to be non-negative,
\begin{equation}
    M_H^{\rm min}(M)
    =
    \frac{M}{
        \mathcal K
        \left(\Phi-\mathcal C_{\rm th}\right)^{\nu_c}
    }.
    \label{eq:minimum_horizon_mass}
\end{equation}
Integrating Eq.~\eqref{eq:full_differential_mass_function} over $d\ln M$ reproduces the total abundance in Eq.~\eqref{eq:total_fpbh}.


\bibliographystyle{JHEP}
\bibliography{bibliography}


\end{document}